\documentclass[aps,showpacs,twocolumn,groupedaddress,prd]{revtex4}

\usepackage[dvips]{epsfig}

\begin{document}

\title{Dilepton low $p_T$ suppression as an evidence of the  Color
  Glass Condensate} 

\author{ M. A. Betemps}
\email{mandrebe@if.ufrgs.br} 
\author{M. B. Gay Ducati}
\email{gay@if.ufrgs.br}

\affiliation{
High Energy Physics Phenomenology Group, GFPAE\\
Instituto de F\'{\i}sica, Universidade Federal do Rio Grande do Sul \\
Caixa Postal 15051, CEP 91501-970, Porto Alegre, RS, Brazil.}

\date{\today}

\begin{abstract}

The dilepton production is investigated in proton-nucleus collisions
in the forward region using the Color Glass Condensate approach.  The
transverse momentum distribution ($p_T$), more precisely the low $p_T$
region where the saturation effects are expected to increase, is
analyzed. The ratio between proton-nucleus and proton-proton
differential cross section for RHIC and LHC energies is evaluated,
showing the effects of saturation at small $p_T$, and presenting a
suppression of the Cronin type peak at moderate $p_T$. These features
indicate the dilepton as a most suitable probe to study the properties
of the saturated regime and the Cronin effect.

\end{abstract}


\maketitle

\section{Introduction}

At high energies, the linear evolution equations, based in the
standard perturbative QCD, predict a high gluon density, requiring
that the growth of the parton density has to have a limit \cite{GLR} -
otherwise violation of the Froissart-Martin bound occurs - and
expected to saturate at a scale $Q_s$, forming a color glass
condensate (CGC)\cite{CGC,RGE1,RGE2,RGE3}.  In this context, the
search for signatures of a CGC description of the saturated regime is
an outstanding aspect of investigation in heavy ion colliders. The
first results of Relativistic Heavy Ion Collider (RHIC), on charged
hadron multiplicity in $Au-Au$ collisions, were treated considering
that the CGC formulation gives a natural qualitative explanation of the
data\cite{Kharz1}. However, there are several issues to be clarified,
before conclude that the dynamical of the partonic system should be
described by a CGC already at RHIC energies. Particularly the current
data are reasonably described by other models based on different
assumptions \cite{Eskola,Armesto}.  However, the charged multiplicity
distribution in pseudo-rapidity for deuteron-gold collision is
estimated within the CGC formalism at the deuteron fragmentation
region, and pointed out as a probable signature of the saturated
regime \cite{Kharz4}. For a review of the CGC signatures see
Refs. \cite{BlaizotG,McLerran:2004fg}.

In order to investigate the high energy limit of the partonic
interactions, the proton-nucleus scattering was proposed as an ideal
experiment to give evidences of the saturation effects described by the CGC
in the proton fragmentation region
\cite{KovcheMu,DimiMc,IancuVen}. Furthermore, the dilepton production
was shown to be a sensitive probe of the perturbative shadowing and saturation
dynamics in proton-proton, proton-nucleus and nucleus-nucleus
scattering
\cite{DYdip,dipoledilep,Raufpp,Rauf2,FQZ,BMS,dileptonGelisJJI,dileptonGelisJJII,dileptonJamal}
in the forward kinematical region. It is an interesting observable
since it is a clean process in which there
is no strong interaction with the nuclear medium final state.

In this work we investigate quantitative features of the dilepton
production in the forward region of proton-nucleus collisions in the
context of the color glass condensate. In particular, the transverse
momentum ($p_T$) distribution is studied focusing attention to the
small $p_T$ region, where the saturation effects are expected to be
more significant.  The main goal of this work is to show the effects
of saturation at small $p_T$ described by the CGC. Their presence at
small $p_T$ should be considered as a possible signature of the
saturation effects when contrasted with proton-proton results. This
comparison is performed evaluating the ratio between proton-nucleus
and proton-proton cross section. This ratio shows two diferent
behaviors; it presents Cronin type peak (if a local Gaussian for the
correlator function is used) and a large suppression (if a non-local
Gaussian is used), being a most suitable probe of the status of the
Cronin effect as a final or initial state effect.  This work is
organized as follows. In the next section one presents a brief
discussion on Color Glass Condensate formalism. In Sec. 3 the dilepton
production cross section within the CGC formalism is presented. The
Sec. 4 is devoted to the study on the color field correlator, which is
a fundamental factor in the CGC approach. The numerical results are
given and discussed in the last section where our conclusions are also
presented.

\section{The Color Glass Condensate}

The Color Glass Condensate picture holds in a frame in which the
hadron propagates at the speed of light and, by Lorentz contraction,
appears as an infinitesimally thin two-dimensional sheet located at
the light cone. The formalism supporting this picture is in terms of a
classical effective theory valid at small $x$ region (large gluon
density), and was originally proposed to describe the
gluons in large nuclei \cite{CGC}.

At small $x$ and/or large $A$ one expects the transition of the regime
described by the standard perturbative QCD
(Dokshitzer-Gribov-Lipatov-Altarelli-Parisi (DGLAP),
Balitsky-Fadin-Kuraev-Lipatov (BFKL)) to a new regime where the
processes like recombination of partons should be important in the
parton cascade \cite{GLR}. In this regime, the growth of the parton
distribution is expected to saturate below a specific scale $Q_s$,
forming a Color Glass Condensate \cite{CGC,RGE1,RGE2,RGE3}.  This
saturated field, meaning the dominant field or gluons, has a large
occupation number and allows the use of semi-classical methods.  These
methods provide the description of the small $x$ gluons as being
radiated from fast moving color sources (parton with higher values of
$x$), being described by a color source density $\rho_a$, with
internal dynamics frozen by Lorentz time dilatation, thus forming a
color glass. The small $x$ gluons saturate at a value of order
$xG(x,Q^2)\sim 1/\alpha_s>>1$ for $Q^2\lesssim Q_s^2$, corresponding
to a multi-particle Bose condensate state. The color fields are driven
by the classical Yang-Mills equation of motion with the sources given
by the large $x$ partons.  The large $x$ partons move nearly at the
speed of light in the positive $z$ direction.

In the CGC approach the light cone variables are employed, where,
$x^{\mu}\equiv (x^+,x^-,x_{\perp})$, with $x^{\pm}\equiv
1/\sqrt{2}(x^0\pm x^3)$ and $x_{\perp}\equiv (x^1,x^2)$, and
$x^{\mu}p_{\mu}=x^+p^-+x^-p^+-x_{\perp}\cdot p_{\perp}$. The variable
$x^+$ is the time light cone, and $p^-$ its variable conjugated
identified with the energy as $p^-=\frac{(m^2
+p_{\perp}^2)}{2p^+}$. The large $x$ partons (fast) have momentum
$p^+$, emitting (or absorbing) soft gluons with momentum $k^+<<p^+$,
generating a color current only with the $+$ component $J_a^{+}=\delta
(x^-)\rho_{a}$. In this framework, the nucleus is situated at $x^-
\approx 0$, with an uncertainty $\Delta x^- \lesssim 1/k^+$, and there
is a separation between fast and soft partons, implying that the
former have large lifetime while soft partons have a short
lifetime. These features assure that the color source density $\rho_a$
should be considered time independent, since for the emitted soft
gluons (small $x$ gluons) the source is frozen in time.  However,
after a time interval of order $1/\varepsilon_p$ ($\varepsilon_p$
being the energy of the source in the light-cone) the configuration of
the source is different. In order to have a
gauge-invariant formulation, the source $\rho_a$ must be treated as a
stochastic variable with zero expectation value. For these
reason, an average over all configurations is required
and it is performed through a weight function $W_{\Lambda^+}[\rho]$, which
depends upon the dynamics of the fast modes, and upon the
intermediate scale $\Lambda^+$, which defines fast ($p^+>\Lambda^+$)
and soft ($p^+<\Lambda^+$) modes.
The classical fields obey the Yang-Mills equation of motion,
\begin{eqnarray}
D_{\nu}F^{\nu\mu}_{a}(x)=\delta^{\mu +}\rho_a(x^-,x_{\perp}),
\end{eqnarray}
and a physical observable is obtained by averaging the solution to this
equation over all configurations of $\rho_a$, with the gauge-invariant
weight function $W_{\Lambda^+}[\rho]$. 

The effective theory is valid only at soft momenta of order
$\Lambda^+$. Indeed, going to a much softer scale, there are large
radiative corrections which invalidate the classical
approximation. The modifications to the effective classical theory is
governed by a functional, nonlinear, evolution equation, originally
derived by Jalilian-Marian, Kovner, Leonidov and Weigert (JIMWLK)
\cite{RGE1,RGE2} for the statistical weight function
$W_{\Lambda^+}[\rho]$ associated with the random variable $\rho_a(x)$.

The solution for such functional evolution equation is not well
determined yet and in order to make predictions or comparison with
data, some phenomenological treatment should be given to the weight
function.  In this work an approximation to the weight function which
is reasonable when we have large nuclei is used and consists in taking
a Gaussian form \cite{IancuVen,Gelisqq1,Gaussian}. As a consequence,
most calculations in the CGC should be done quasi-analytically. In
Ref. \cite{Gaussian} it is shown that a Gaussian weight function can
accommodate both the BFKL evolution of the gluon distribution at high
transverse momenta, and the gluon saturation phenomenon at low
transverse momenta.  A non-local Gaussian distribution of color
sources has been predicted in Ref. \cite{Iancu:2001ad} as a mean-field
asymptotic solution for the JIMWLK equation and provides some
modifications concerning phenomenological properties of the
observables \cite{BlaizotGVI}. The local Gaussian weight function
assures that the color sources are correlated locally, on the other
side, the non-local Gaussian allows correlations over large distances.
In the following sections the phenomenological consequences of the
choice of a local or non-local Gaussian type for the weight function
are investigated concerning the transverse momentum of the dileptons.

\section{Dilepton production in the CGC approach}

\begin{figure}[ht]
\scalebox{0.95}{\includegraphics{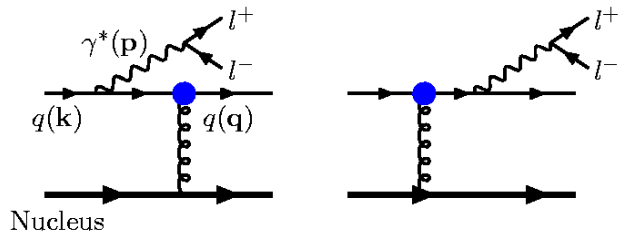}}
\caption{Dilepton production in the CGC.}
\label{photonfig1}
\end{figure}

At high energies, the dilepton production in hadronic collisions looks
like a bremsstrahlung of a virtual photon with momentum ${\bf p}$
decaying into a lepton pair, which can occur before and after the
interaction of the quark (momentum ${\bf k}$) with the dense saturated
gluonic field (momentum ${\bf q}$) of the target, in our case the
nucleus $A$.  We consider only the diagrams where the photon emission
occurs before and after before the interaction with the nucleus, since
the emission considering before and after the interaction vanishes
\cite{Gelis:2002ki}. The dilepton production can be summarized in
Figure \ref{photonfig1} \cite{dipoledilep,Raufpp,dileptonGelisJJI},

Considering the Fig. \ref{photonfig1}, the differential cross section
for the dilepton production in the CGC approach, for a collinear quark
($k_T=0$), is written as \cite{dileptonGelisJJI},
\begin{widetext}
\begin{eqnarray}\nonumber
\frac{d\sigma_{incl}^{qA\rightarrow
    ql^+l^-X}}{dzd^2{p_T}d \log
    M^2}&=&\pi R^2_A f_{q}^2\frac{2\alpha_{em}^2}{3\pi}\int \frac{d^2 
    l_T}{(2\pi)^4} C(
    l_T)
\left\{\left[\frac{1+(1-z)^2}{z}\right]\right.\\\nonumber
&\times& \frac{z^2 l_T^2}{[ p_T^2+M^2(1-z)][( p_T-z
    l_T)^2+M^2(1-z)]}\\
&\!\!-&\!\!\left.z(1-z)M^2\left[\frac{1}{[ p_T^2+M^2(1-z)]}-\frac{1}{[(
    p_T-z l_T)^2+M^2(1-z)]}\right]^2 \right\},
\label{cs1}
\end{eqnarray}
\end{widetext}
where $f_q$ represents the fraction of the electron charge carried by
the quark $q$. The squared quark charge is $e^2_q=f^2_q e^2$ and the
charge $e^2 $ from $e_q^2$ was incorporated in the
$\alpha_{em}^2$ in the expression. $R_A$ is the nuclear radius,
$z\equiv p^-/k^-$ is the energy fraction of the proton carried by the
virtual photon, 
$p_T$ and $M^2$ are the transverse momentum and  the squared
invariant mass of the lepton pair, respectively;
$l_T=q_T+p_T$ is the total transverse momentum 
transfer between the nucleus and the quark. 
The function $C(l_T)$ is the field correlator function and defined by
\cite{Gelisqq1}, 
\begin{eqnarray}
C(l_T)\equiv \int d^2 x_{\perp} e^{il_T\cdot
  x_{\perp}}\langle U(0)U^{\dagger}(x_{\perp})\rangle _{\rho},
\label{defCk}
\end{eqnarray}
with the averaged term representing the average over all
configurations of the color fields in the nucleus, $U(x_{\perp})$ is a
matrix in the $SU(N)$ fundamental representation which represents the
interactions of the quark with the classical color field. The
correlator considers the two diagrams, being the interaction at two
transverse locations, and all the information about the nature of the
medium crossed by the quark is contained in the function $C(l_T)$. In
particular, it determines the dependence on the saturation scale $Q_s$
(and on energy).

In order to obtain a hadronic cross section, the validity of
the collinear factorization in the fragmentation region is assumed and
the expression in Eq. (\ref{cs1}) is convoluted with the partonic
distribution function in 
the proton (deuteron or nucleus), as was done in
\cite{Raufpp,dileptonJamal} and the cross section reads as,
\begin{eqnarray}\nonumber
\frac{d\sigma^{pA\rightarrow
    ql^+l^-X}}{dp_T^2\,dM\,dx_F}&\!\!=&\!\!\frac{4\pi^2}{M} R^2_A 
\frac{\alpha_{em}^2}{3\pi}\frac{1}{x_1+x_2}\\
\times \int \frac{dl_T}{(2\pi)^3}&\!\!\!\!\! & \!\!\!\!\! l_T
\,    W(p_T,l_T,x_1)\,C(l_T,x_2,A),
\label{eqcsh}
\end{eqnarray}
where $x_F$ is the longitudinal momentum fraction given by
$x_F=x_1-x_2$, and $x_1$ and $x_2$ are the momentum fraction carried
by the quark from the proton and by the gluonic field from the
nucleus, respectively. The expression (\ref{eqcsh}) is valid in the
forward region, which means positive $x_F$, or positive rapidities
$y$, and the variables $x_1$ and $x_2$ are defined by,
\begin{eqnarray}
x_{{1 \choose 2}}&=&\frac{1}{2}\left\{\sqrt{x_F+4\frac{M_T^2}{s}}
(\pm)\,  x_F\right\},
\end{eqnarray}
or
\begin{eqnarray}
x_{{1 \choose 2}}&=&\sqrt{\frac{M^2_T}{s}}e^{\pm y},
\label{x1x2}
\end{eqnarray}
where $M_T^2=M^2+p_T^2$ is the squared dilepton transverse mass and $s$ is the
squared center of mass energy. 
Here, using the structure function $F_2(x,Q^2)=\sum_i
e_{q_i}^2x[q_i(x,Q^2)+\bar{q}_i(x,Q^2)]$, the weight function
$W(p_T,l_T,x_1)$ can be written as,
\begin{eqnarray}\nonumber
&&W(p_T,l_T,x_1)=\int_{x_1}^{1}dz\,z  F_2 (x_1/z,M^2)\\\nonumber
&&\times \left\{
 \frac{(1+(1-z)^2)z^2 l_T^2}{[ p_T^2+M^2(1-z)][( p_T-z
    l_T)^2+M^2(1-z)]}\right. \\\nonumber
&\!\!-&\!\!z(1-z)M^2\left[\frac{1}{[ p_T^2+M^2(1-z)]}\right.\\
&\!\!-&\!\! \left.\left.\frac{1}{[(
    p_T-z l_T)^2+M^2(1-z)]}\right]^2 \right\}.
\label{cs2}
\end{eqnarray}
In our calculations the CTEQ6L parametrization \cite{Cteq6} was used
for the structure function, and the lepton pair mass gives the scale
for the projectile quark distribution. The function $W(p_T,l_T,x_1)$
plays the role of a weight function, selecting the regions of
dominance on $l_T$ contributing to the cross section.

In Eq. (\ref{eqcsh}) the correlator function appears with an energy
dependence (dependence on $x_2$), which is not included in the
original McLerran-Venugopalan model. One includes such dependence in
the field correlator function only in the saturation scale
$Q_{s,A}\rightarrow Q_{s,A}(x)$ to simulate a low $x$ evolution, as
was done in the Ref. \cite{BlaizotGVI}, in order to investigate the
effects of the $x$ evolution in the dilepton $p_T$ spectra. The $x$
dependence is parametrized in the form proposed by Golec-Biernat and
W\"usthoff (GBW) \cite{GBW} ($Q_s^2=(x_0/x)^{\lambda}$), with the
parameters taken from the dipole cross section extracted from the fit
procedure by GBW \cite{GBW} and CGCfit \cite{IIMunier}
parametrizations, which will be discussed later.

\begin{figure}[ht]
\begin{tabular}{c}
\scalebox{0.45}{\includegraphics{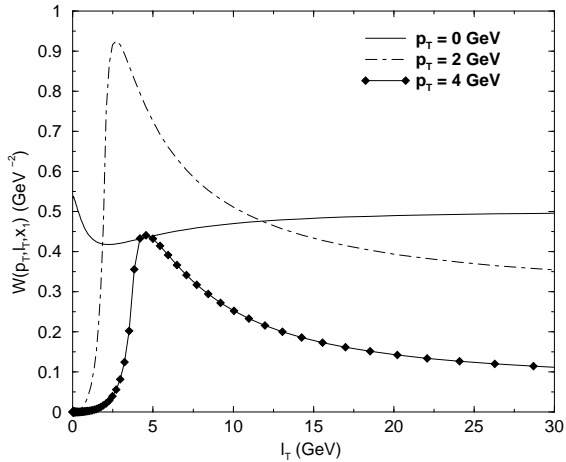}}
\end{tabular}
\caption{Weight function for lepton pair mass $M=3$ GeV and rapidity
  $y=2.2$ versus $l_T$.} 
\label{weight}
\end{figure}

 In Fig. \ref{weight}, we plot the weight function for an
specified lepton pair mass $M=3$ GeV, in the forward region (positive
$x_F$), with a positive value of the rapidity $y=2.2$, considering the
center of mass energy $\sqrt{s}=350$ GeV (RHIC). In such figure the
results for three representative values of $p_T$ are presented, where
a peak at $l_T \approx p_T$ and a suppression at $l_T<p_T$ are in
order. These characteristics assure that the weight function selects
the values of $l_T$ larger than $p_T$.  Moreover, one verifies that
larger values of $p_T$ provide a reduction in the normalization of
the weight function at large values of $l_T$, when compared with the
normalization at $p_T=0$ GeV.  This $p_T$ behavior of the weight
function is essential in order to determine the spectrum. As
will be verified in the Sec. \ref{results}, the $p_T$
distribution is suppressed for large values.

All high density effects on the nucleus are encoded in the 
field correlator function. It is well known that the saturation
effects in the correlator function $C(l_T,x,A)$ are present below the
saturation scale, meaning the small $l_T$ region (In the next section
one observes this feature in the Fig. \ref{funcCk}). Such behavior
determines that only at small $p_T$ the effects of saturation in the
function $C(l_T,x,A)$ should be measurable, once the weight function
selects the values of $l_T$ larger than $p_T$.

To make a quantitative prediction to the dilepton production, the
correlator function $C(l_T,x,A)$ has to be determined.  It plays an
important role in the Color Glass Condensate formalism and should be
compared with the dipole cross section. It is related to dipole
Fourier transform and in the following, the function $C(l_T,x_2,A)$
will be discussed, through the analyses of some phenomenological
models.

\section{The color field correlator $C(l_T,x,A)$}
\label{funcCkcap}

The function $C(l_T)$ is considered a fundamental quantity in the CGC
formalism, since it contains all the information on high density
effects. It can be related to the
Fourier transform (F.T.) of the dipole cross section in the following
way \cite{dileptonGelisJJII,BGBK,VPGMVTMIII},
\begin{eqnarray}
C(l_T)\!=\!\frac{1}{\sigma_0}\!\! \int d^2 x_{\perp} e^{il_T\cdot
  x_{\perp}}[\sigma_{dip}(x_{\perp}\rightarrow
  \infty)-\sigma_{dip}(x_{\perp})], 
\label{originCk}
\end{eqnarray}
where $\sigma_0$ is the normalization of the dipole cross section at
the saturation region ($x_{\perp}\rightarrow \infty$).
Considering the GBW model for the dipole cross section
$\sigma_{dip}(x_{\perp},x)=\sigma_0[1-\exp(-Q_s^2(x)x_{\perp}^2/4)]$ \cite{GBW}
the correlator function can be 
written as \cite{dileptonGelisJJII,VPGMVTMIII},  
\begin{eqnarray}
C(l_T,x,A)_{GBW}=\frac{4\pi}{Q_s^2(x,A)}e^{-\frac{l_T^2}{Q_s^2(x,A)}},
\label{GBWFT}
\end{eqnarray}
where a simple dependence on energy ($x$) and atomic number ($A$) was
taken into the saturation scale. Namely, the nuclear saturation scale
was considered as $Q_s^2(x,A)=A^{1/3}Q_s^2(x)$ with $Q_s^2(x)$ being
the proton saturation scale parametrized of the form proposed by GBW
$Q_s^2=(x_0/x)^{\lambda}$ \cite{GBW}, where the parameters
$x_0=3.10^{-4}$ and $\lambda=0.288$ were determined from the fit to
the Hadron Electron Ring Accelerator (HERA) data.  This ansatz to the
nuclear dependence of the saturation scale was studied in the
Ref. \cite{FreundPRL} concerning the $eA$ data, and was shown to be a
consistent approximation for large nuclei and moderate energies.

\begin{figure}[ht]
\scalebox{0.45}{\includegraphics{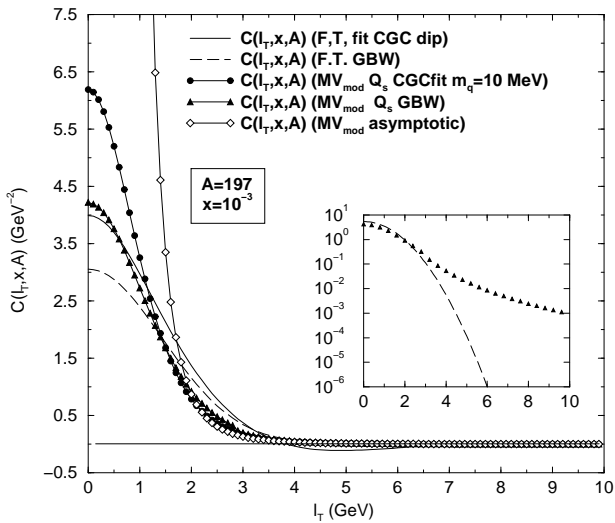}}
\caption{Correlator $C(l_T,x,A)$ as a function of  $l_T$.}
\label{funcCk}
\end{figure}

However, the GBW F.T. (Eq.  (\ref{GBWFT})) does not recover the perturbative
behavior at large $p_T$, since it presents an exponential tail, as we
show in the Fig. \ref{funcCk}.

Considering the McLerran-Venugopalan (MV) model, the function $C(l_T)$
has no energy dependence and should be computed by taking the MV
dipole cross section \cite{dileptonGelisJJII}, 
\begin{equation}
\sigma_{dipole}(r_{\perp})=\pi
R^2\! \left[1-e^{\left(-\frac{Q_s^2}{\pi}\int\frac{dp}{p^3}
(1- J_{0}(p r_{\perp}))\right)}\right].
\end{equation}
The Fourier transform can be numerically computed in the form \cite{Gelisqq1},
\begin{eqnarray}
C(l_T)_{MV} \equiv \int d^2 x_{\perp} e^{il_T\cdot
  x_{\perp}}e^{-\frac{Q_s^2}{\pi}\int\frac{dp}{p^3}
\left(1-J_0(px_{\perp})\right)},
\label{funcCkMV}
\end{eqnarray}
where the value of $Q_s^2$ is fixed depending on the energy. However,
no $x$ evolution is presented in the MV model and the energy
dependence is introduced in the correlator by fixing the value of the
saturation scale.  Following the equation (\ref{funcCkMV}), we propose
to introduce the dependence on the energy and nuclei into the
saturation scale in the form,
\begin{eqnarray}\nonumber
C_{MV_{mod}}(l_T,x,A) & = & \int d^2 x_{\perp} e^{il_T\cdot
  x_{\perp}}\\
& \times & e^{-\frac{Q_s^2(x,A)}{\pi}\int\frac{dp}{p^3}
\left(1-J_0(px_{\perp})\right)}.
\end{eqnarray}
The nuclear saturation scale is parametrized in the form presented
previously, where the $x$ dependence in the saturation scale
$Q_s^2(x)$ is considered from the parameters extracted from the fit to
HERA data and should be taken from GBW saturation model \cite{GBW}, or
from a dipole cross section based on the CGC approach \cite{IIMunier}.
Here it should be interesting to point out that in the recent work of
Ref. \cite{BlaizotGVI}, the Cronin effect was studied in the MV model,
and the same energy dependence for the saturation scale was
considered. Such a simple inclusion of the quantum corrections results
in a disagreement with the RHIC $d-Au$ data at forward rapidities
concerning the Cronin effect \cite{BRAHMSQM,Arsene}. In that work
\cite{BlaizotGVI}, a non-local Gaussian distribution for $\langle
U^{\dagger}(0)U(x_{\perp})\rangle$ was introduced and the shape of the
curves agrees with the Broad Range Hadron Magnetic Spectrometer
(BRAHMS) data at large rapidities, however presents large suppression
at central rapidity.  This disagreement shows that the dynamics of the
CGC is a subject which deserves more comprehensive studies.  Here we
point out that the dilepton transverse momentum analyzed with local
and non-local Gaussian correlator is a good observable to investigated
such dynamics.

In the large $l_T$ ($l_T>>Q_s$) limit, the correlator function should
recover the perturbative behavior ($1/l_T^4$), and considering the MV
model the correlator function can be expanded and written in a simple
analytic expression \cite{Gelisqq1},
\begin{eqnarray}\nonumber
&&C_{MV_{mod}}(l_T,x,A)|_{l_T>>Q_s}
=2\frac{Q_s^2(x,A)}{l_T^4}\\
&\times & \left(1+\frac{4Q_s^2(x,A)}{\pi l_T^2}
\left[\ln\left(\frac{l_T}{\Lambda_{QCD}}\right)-1\right]\right),
\end{eqnarray}
which emphasizes the large saturation effects in
the small $l_T$ region, as in the Fig. (\ref{funcCk}).

In a recent work, Ref. \cite{IIMunier} it has been analyzed the
structure function $F_2(x,Q^2)$ for $x<10^{-2}$ and $0.045\leq Q^2
\leq 45 $ GeV$^2$, within the dipole picture, taking an expression to
the dipole cross section which interpolates the BFKL solution at
$r<<1/Q_s(x)$ and the saturated behavior at $r>>1/Q_s(x)$, where the
scattering amplitude saturates at one. The parametrized dipole cross
section can be written as $\sigma_{dip}(x,r)=2\pi R^2 {\cal
N}(rQ_s,x)$ with \cite{IIMunier},
\begin{eqnarray}
\begin{array}{lll}
{\cal N}(rQ_s,Y)={\cal
  N}_0\left(\frac{rQ_s}{2}\right)^{2\left(\gamma_s+
\frac{\ln(2/rQ_s)}{\kappa\lambda Y}\right)} &\hbox{to
  }& rQ_s\leq 2\\ 
{\cal N}(rQ_s,Y)=1-e^{-a\ln^2(brQ_s)} & \hbox{to
  }& rQ_s\geq 2,
\end{array}
\label{dipCGChep}
\end{eqnarray}
where $Y=\ln(1/x)$. There are three free parameters:
the proton radius $R$, the value $x_0$ of $x$ at which the saturation
scale has to be equal to 1,  and the parameter which controls the
energy dependence of the saturation scale $\lambda$. 
The parameters $a$ and $b$ are determined to assure that ${\cal N}$ is
continuous at $rQ_s=2$ (at least at first derivative). From the fit to
the HERA data for the inclusive structure function $F_2(x,Q^2)$ the
parameters depend on the quark mass $m_q$.  

Following this dipole cross section, we construct a function
$C(l_T,x_2,A)$ taking the Eq. (\ref{originCk}), that will be called
CGCfit, and obtain the following expression,
\begin{widetext}

\begin{eqnarray}\nonumber
C(l_T,x,A)_{CGC}&=&2\pi\left(\int_{0}^{2/Q_{s}}rdrJ_{0}(l_Tr)\left(1-{\cal
    N}_{0}\exp\left\{ 2\ln\left(\frac{rQ_{s}}{2}\right)
\left[\gamma_{s}+\frac{\ln(2/rQ_{s})}{\kappa\lambda\ln1/x}\right]\right\}
\right)\right.\\
&+&\left.\int_{2/Q_{s}}^{\infty}rdrJ_{0}(l_Tr)e^{-a\ln^{2}(brQ_{s})}\right).
\end{eqnarray}  

\end{widetext}
The energy and nuclear dependences are introduced with
$Q_s^2(x,A)=A^{1/3}\left(\frac{x_0}{x}\right)^{\lambda}$.

Considering the two models for the dipole cross section (GBW and
CGCfit) there is a set of parameters which determine the saturation
scale, where the ones used in this work are presented in the Table
\ref{param} ( the set of parameters are identified as fit1, fit2 and
fit3), where the value of the saturation scale was calculated at
$x=10^{-3}$ for gold. It is shown that the recent CGC fit
parameters provide a small value for the saturation scale, and this
behavior should affect the dilepton production, as we will see in the
next section. The nuclear radius is taken from the Woods-Saxon
parametrization of the form, $R_A=1.2 A^{1/3} fm$, while the proton
radius is taken from the fits and presented in the Table \ref{param}.
\begin{widetext}

\begin{table}[h]
\begin{center}
\begin{tabular}{c||c|c|c}
Parameter & GBW & CGC fit $m_q=10$ MeV  & CGC fit $m_q=140$ MeV 
\\
& (fit1) & (fit2) &(fit3)\\
\hline
$x_0$ & $3\times 10^{-4}$ & $1.06\times 10^{-4}$ & $0.267\times
10^{-4}$\\
\hline
$\lambda$ & 0.288 & 0.285 &0.253 \\
\hline
$Q_s^2$ ($x=10^{-3}$, $A=197$) & 4.114 GeV$^{2}$ & 3.069 GeV$^2$ &
2.327 GeV$^2$\\
\hline
$R_p$ (Proton radius) & 0.6055 fm & 0.566 fm & 0.641 fm \\
\end{tabular}
\end{center}
\caption{Parameters of saturation scale from GBW and CGCfit.}
\label{param}
\end{table}

\end{widetext}

In the Figure \ref{funcCk} the function $C(l_T,x_2,A)$ is shown for
the models discussed here and one verifies the large saturation
effects at small $l_T$ when we compare the functions obtained from the GBW
and CGCfit with the asymptotic behavior of the correlator function. At
this point it is interesting to emphasize some features of the
functions $C(l_T,x_2,A)$ extracted from the Fourier transform of GBW
dipole cross section and the Fourier transform of the
CGCfit. Considering the GBW F.T.  (dashed-line) the function
$C(l_T,x_2,A)_{GBW}$ depends on $e^{-l^2_T}$ and is suppressed at
large $l_T$ (this behavior is shown in the inner plot on
Fig. \ref{funcCk}).  It results in an unrealistic suppression of the
observable cross section at large $p_T$, as emphasized in the
Refs. \cite{dileptonGelisJJII,VPGMVTMIII}.  Considering the results
from CGC fit (solid-line), the function $C(l_T,x_2,A)$ presents
negative values at moderate $l_T$ (as can be seen in the Figure
\ref{funcCk}) and this behavior should be due to the continuity only
at first derivative or by the approximations in the construction of
the dipole cross section model \cite{IIMunier}. Having those aspects
in mind, the function $C(l_T,x_2,A)$ based on the McLerran-Venugopalan
model, including an energy and nuclear dependence in the saturation
scale, is employed here. The parameters to the saturation scale are
taken from the fit1 (triangle-up-line) and fit2 (circle-line) and one
verifies in the Fig. \ref{funcCk} that the value of the saturation scale
provides a small difference in the correlator obtained from the MV$_{mod}$
model at small $l_T$.

The correlator function is suppressed at large $l_T$, and the weight
function suppress the values of $l_T$ smaller than $p_T$, the behavior
of the cross section coming from the balance between these two
quantities.  In such balance, the small $p_T$ dileptons clearly are
the dominant contribution for the cross section and provide a physical
probe for the CGC and for the models to the color field correlator
function.

\begin{figure}[h]
\scalebox{0.45}{\includegraphics{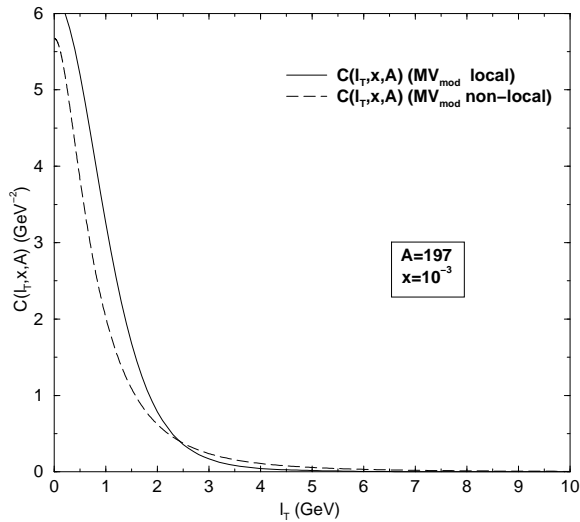}}
\caption{Field correlator function with local and non-local Gaussian
  weight function.}
\label{Cknon}
\end{figure}

The field correlator function presented up to here in MV model is
obtained considering a local Gaussian function for the weight function
$W_{\Lambda^+}[\rho]$. The correlator function is defined by
\begin{eqnarray}
C(l_T)\equiv \int d^2 x_{\perp} e^{il_T\cdot
  x_{\perp}}\langle U(0)U^{\dagger}(x_{\perp})\rangle _{\rho},
\label{defCk2}
\end{eqnarray}
and the local Gaussian enters in the calculation of the averaged term
$\langle U(0)U^{\dagger}(x_{\perp})\rangle _{\rho}$. The consideration
of a non-local Gaussian function modifies the correlator in such way
that it is written as \cite{Gaussian,BlaizotGVI}
\begin{eqnarray}
C(l_T,x,A) \equiv  \int d^2 x_{\perp} e^{il_T\cdot
  x_{\perp}}e^{\chi(x,x_{\perp},A)},  
\end{eqnarray}
with 
\begin{eqnarray}\nonumber
\chi(x,x_{\perp},A)&\equiv& -\frac{2}{\gamma c} \int
  \frac{dp}{p}(1-J_0(x_{\perp}p))\\
\times &&\ln\left(1+\left(\frac{Q_{2}^2(x,A)}{p^2}
\right)^{\gamma}\right),
\end{eqnarray}
where, $\gamma$ is the anomalous dimension ($\gamma\approx$ 0.64 for
BFKL) and $c\approx $ 4.84 \cite{Gaussian,BlaizotGVI}.  This non-local field
correlator function is plotted in the Figure \ref{Cknon} in contrast
with the same correlator obtained with the local Gaussian weight
functional.

The physical effect of the nonlocal Gaussian weight function is that
the gluon sources are no longer correlated locally, as in the
local Gaussian, but correlate over larger distances. This implies in a
more drastic reduction of the gluon density, as can be seen at small
$l_T$ in the Fig. \ref{Cknon}, where the solid line represents the
correlator function with a local Gaussian weight function and the
long-dashed line represents the non-local Gaussian weight function.
The effect of the local or non-local Gaussian weight function in the
$p_T$ dilepton spectra will be discussed in the next section in the
context of the defined ratio $R_{pA}$.

Having addressed all the fundamental aspects to develop the
calculation of the dilepton transverse momentum in the CGC formalism,
one presents in the next section the numerical predictions using such
approach and the discussions.

\section{Results and Discussions}
\label{results}

In what follows, the numerical results on the dilepton transverse
momentum distribution in CGC are addressed and discussed. We consider
$pA$ collisions at RHIC ($\sqrt{s}=350$ GeV) and LHC energies
($\sqrt{s}=8.8$ TeV) in the proton fragmentation region (positive
rapidities).  The calculations are performed fixing values of
rapidities (or $x_F$) and lepton pair mass $M$.  We use the function
$C(l_T,x_2)$ based on the McLerran-Venugopalan model, however an $x$
dependence through the saturation scale is introduced, taking the
parameters from the HERA data fit procedure GBW (fit1) \cite{GBW} and
CGCfit (fit2) \cite{IIMunier}.  For sake of comparison, the same
differential cross section using the original McLerran-Venugopalan
model, fixing a value to the saturation scale is evaluated.

\begin{figure}[ht]
\scalebox{0.45}{\includegraphics{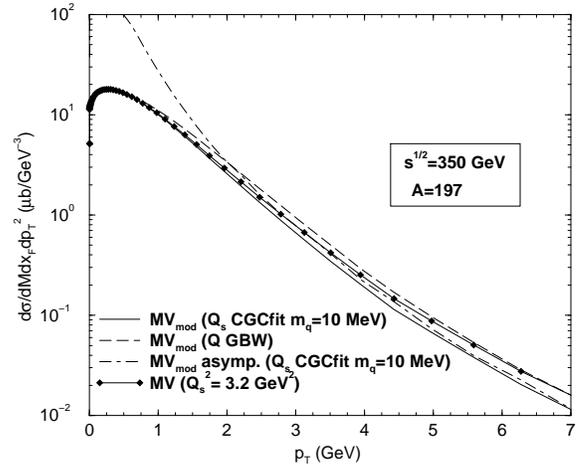}}
\caption{Dilepton production at RHIC energies ($\sqrt{s}=350$ GeV)  in
$pA$ collisions, considering rapidity $y=2.2$ and lepton pair mass $M=3$ GeV.} 
\label{dileptRHIC} 
\end{figure}

\begin{figure}[ht]
\scalebox{0.45}{\includegraphics{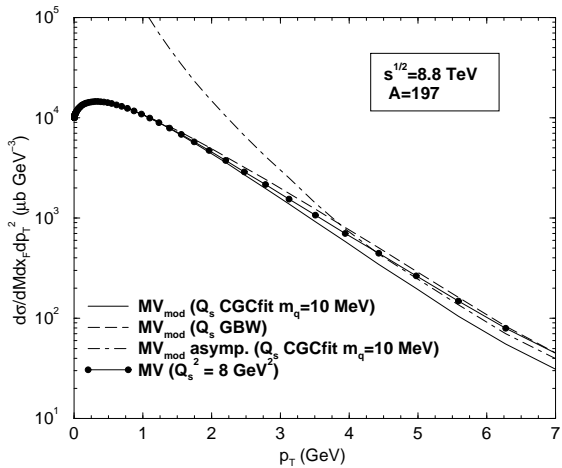}}
\caption{Dilepton production at LHC energies ($\sqrt{s}=8.8$ TeV) in
  $pA$ collisions, considering rapidity $y=2.2$ and lepton pair mass
  $M=3$ GeV.}
\label{dileptLHC}
\end{figure}

In Fig. \ref{dileptRHIC} we present the transverse momentum
distribution for RHIC energies ($\sqrt{s}=350$ GeV) in $pA$
collisions, for a lepton pair mass $M=3$ GeV and as in the
Ref. \cite{dileptonJamal} at rapidity $y=2.2$.  The proton structure
function is taken from the CTEQ6L parametrization \cite{Cteq6}. The
solid line is the calculation with the McLerran-Venugopalan model,
with the $x$ dependence on the saturation scale, taking the parameters
from the fit2; the dashed-line is the same calculation with the saturation
scale taken from the fit1 and the dot-dashed line is the
calculation with the asymptotic behavior of the MV correlator
function.  Considering the transverse momentum distribution at fixed
mass and rapidities, the effects of quantum evolution are not too
relevant in the range of transverse momentum investigated here, once
the parametrization of the saturation scale assures that it
is almost fixed in the region treated in this case, changing only
weakly with the transverse momentum
($x_{2}=\sqrt{\frac{M^2+p_T^2}{s}}e^{-y}$).  Such behavior can be seen
in the Figs. \ref{dileptRHIC} and \ref{dileptLHC}, where the
line-diamond represents the calculation with the MV model, fixing the
saturation scale at a value $Q_s^2=3.2$ GeV $^2$ and $Q_s^2=8$
GeV$^2$, respectively. The $x$ evolution provides a small suppression
of the large $p_T$ dilepton, in both cases.
 
In the Figure \ref{dileptRHIC} the large saturation effects presented
at $p_T \lesssim 2$ GeV are verified if one compares the asymptotic
behavior of the correlator function with the MV$_{mod}$ prediction.
As was shown in the last section, the asymptotic behavior of the
correlator function ($l_T>>Q_s$) depends on the $Q_s^2/l_T^4$, then an
increase of the saturation scale provides an increase in the
differential cross section at large $p_T$, as can be seen in the
Figure \ref{dileptRHIC}.  As a most interesting feature, only at large
$p_T$ the effects of the choice of saturation scale affect the cross
section, and the difference between the predictions being a factor of
2 considering the smallest value of the saturation scale, which is
taken from the fit3, in contrast with the GBW ones.

In Figure \ref{dileptLHC} the dilepton transverse momentum
distribution at LHC energies is shown, taking the same value of
rapidity $y=2.2$ to assure the forward region and to make a comparison
with RHIC energies. The same behavior concerning the saturation
effects is verified, although such effects start to be significant for
$p_T\lesssim 4$ GeV.  The estimative with the MV model was performed
and the suppression at large $p_T$ when the energy dependence is
introduced in the saturation scale can be seen.  The $p_T$ spectra is
enlarged at large $p_T$ if the saturation scale is large, as was verified
for RHIC energies.

In order to avoid any ambiguity with normalization, the ratio between the 
proton-nucleus and proton-proton differential cross section for RHIC
and LHC is defined,
\begin{eqnarray}
R_{pA}=
\frac{\frac{d\sigma(pA)}{\pi R_A^2dMdx_Fdp_t^2}}{A^{1/3}
\frac{d\sigma(pp)}{\pi R_p^2dMdx_Fdp_t^2}}.
\end{eqnarray}
Some attention should be given to the uncertainty in the determination
of the nuclear radius, then each cross section is divided by the
nuclear or proton radius.  The factor $A^{1/3}$ was used in the
denominator to guarantee a ratio $R_{pA}$ about 1 at large $p_T$.

The expression to the ratio $R_{pA}$ in the dilepton
production defined here should be written of the form,
\begin{eqnarray}\nonumber
R_{pA}&&\!\! (y,p_T)=\\
 & &\!\!\!\!\frac{\int d^2 l_T W(p_T,l_T,x_1)C_A(l_T,x_2,A)}
 {A^{1/3}\int d^2 l_T W(p_T,l_T,x_1)C_p(l_T,x_2)},
\label{rpac}
\end{eqnarray}
where $C_A$ is the nuclear correlator function and $C_p$ is the proton
correlator function. The ratio in the Eq. (\ref{rpac}) is similar to
the one obtained in the Ref. \cite{BlaizotGVI} to investigate the
Cronin effect (Eq. (113) in the Ref. \cite{BlaizotGVI}).

The Cronin effect was discovered in the late's 70s \cite{origCron} and
is related to the enhancement of the hadron transverse momentum
spectra at moderated $p_T$ (2-5 GeV) in comparison with the
proton-proton collisions (the ratio between $pA$ and $pp$ present a
peak at moderate $p_T$).  The effect should be interpreted as being
originated by the multiple scatterings of the partons from the proton
propagating through the nucleus, resulting in a broadening of the
transverse momentum of the initial partons. This indicates the
Cronin effect as an initial state effect.  The Cronin effect was
measured by the RHIC experiments, in $Au-Au$ and $d-Au$ collisions,
however, the theoretical approaches cannot describe the effect in all
the range of rapidity measured by the collaborations
\cite{Accardi:2004fi}.  Although the Cronin effect concerns the hadron
transverse momentum spectra, it is also expected  in the
dilepton transverse momentum spectra, since the effect of multiple
scatterings is an initial state effect. Moreover, the ratio obtained in
this work is similear to that is used to investigate the Cronin
effect \cite{BlaizotGVI}.

\begin{figure}[ht]
\scalebox{0.5}{\includegraphics{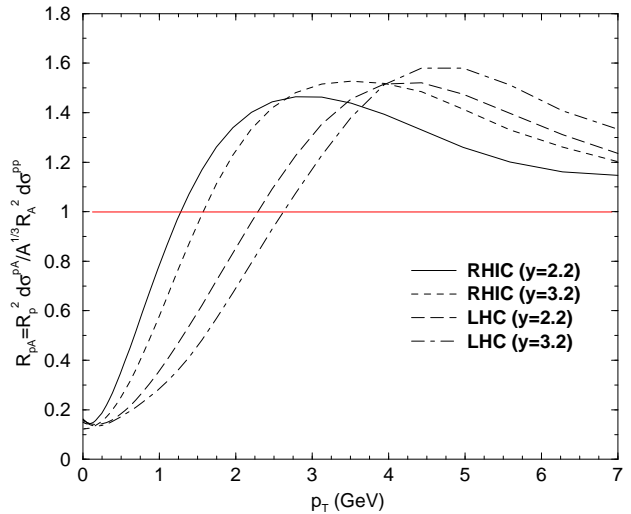}}
\caption{Ratio between proton-nucleus and  proton-proton at RHIC and
  LHC energies by the CGC approach at distincts rapidities with
  local Gaussian distribution for the weight function $W[\rho]$.}
\label{ratio}
\end{figure}

In the Fig. (\ref{ratio}) one presents the results for the ratio
$R_{pA}$ to RHIC and LHC energies considering a correlator field
function $C(l_{T},x,A)$ obtained from a local Gaussian distribution
for the weight function $W_{\Lambda^+}[\rho]$. For RHIC energies the
solid line represents the calculation for rapidity $y=2.2$ and the
dashed line for rapidity $y=3.2$. For LHC energies the long-dashed
line represents the calculation for rapidity $y=2.2$ and the
dot-dashed line for rapidity $y=3.2$. It is verified that at moderate
$p_T$ the calculations show a Cronin type peak for RHIC and for LHC
(there is a suppression of the dilepton production at RHIC and LHC
energies comparing with the proton-proton collisions at small $p_T$,
at intermediate $p_T$ the ratio is larger than 1, and there is a
suppression at large $p_T$ up to get the value 1). Such peak increases
and is shifted to larger $p_T$ at larger rapidities, due to the fact
that the saturation scale grows with the rapidity and no evolution is
taking into account.

Concerning the Cronin effect, such peak is enlarged at large
rapidities if the local Gaussian correlator function with the same
energy dependence implemented here is used \cite{BlaizotGVI}, in
complete disagreement with the BRAHMS experiment at forward rapidities
\cite{BRAHMSQM,Arsene}. In the same Ref. \cite{BlaizotGVI} the Cronin effect
was studied using a non-local Gaussian distribution for the weight
function and the Cronin peak suppression is reached. However, there is a
suppression on the normalization at centrality region, which is not
consistent with the RHIC data in central rapidity \cite{BRAHMSQM,Arsene},
emphasizing that the the non-local Gaussian weight function
should be the right physics at forward rapidities.

\begin{figure}[ht]
\scalebox{0.5}{\includegraphics{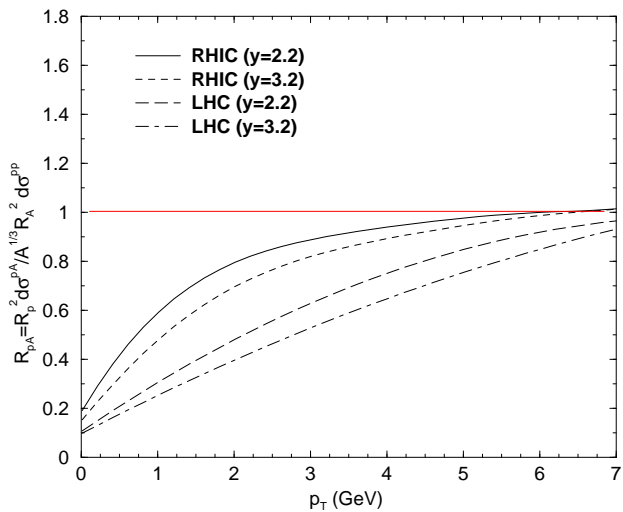}}
\caption{Ratio between proton-nucleus and  proton-proton at RHIC and
  LHC energies by the CGC approach at distincts rapidities with
  non-local Gaussian distribution for the weight function $W[\rho]$.}
\label{rationonloc}
\end{figure}

On the dilepton side, the behavior of the ratio $R_{pA}$, shows the
same features of the Cronin peak at forward rapidities when
investigated with the local Gaussian (the peak is enlarged and shifted
to high $p_T$ at large rapidities). However, the ratio $R_{pA}$ was
also investigated with the correlator function obtained from a
non-local Gaussian weight function $W_{\Lambda^+}[\rho]$, and is
presented in the Fig. \ref{rationonloc}.  The suppression of the ratio
$R_{pA}$ is verified showing exactly the same features presented in
the Cronin effect \cite{BlaizotGVI}, being a possible clean observable
to study this property. Although, the Cronin effect was considered as
a final state one in the Ref. \cite{CroninFS}, in our analysis, the
dilepton production seems to clarify this aspect. It was obtained that
the Cronin type peak (or the suppression of the Cronin peak) in the
dilepton $p_T$ distribution appears as an initial state effect. In the
Fig. (\ref{rationonloc}) the solid line represents the calculation for
rapidity $y=2.2$ and the dashed line for rapidity $y=3.2$ at RHIC
energies. For LHC energies the long-dashed line represents the
calculation for rapidity $y=2.2$ and the dot-dashed line for rapidity
$y=3.2$.

At RHIC energies, the effect of suppression appears if the nonlocal
Gaussian in the correlator color field is used, sugesting the
measurement of such suppression, although the detectors should not be
able to measure such behavior \cite{RHICdetect}. On the other hand, at
LHC energies, the suppression of the ratio $R_pA$ reach large values
of $p_T$ and such suppression increases with the rapidity. It
is interesting to address that at LHC the experimental facilities
provide a detection of dileptons in the forward region with transverse
momentum above 1.5 GeV, depending on the rate of the signal of the
observable and on the signal from physics and machine sources
\cite{Karel}. This feature assures that at LHC energies such
suppression behavior should be detectable too.

\section{Conclusions}

In this work the large saturation effects described by the Color Glass
Condensate in the dilepton production at small $p_T$ region and the
dependence of the large $p_T$ spectra on the saturation scale value
are verified.  Although at RHIC, the transverse momentum distribution
of the dilepton should not be measurable in the very small $p_T$, at
intermediate $p_T$, the comparison between $pp$ and $pA$ cross
section, provides a tool to study the Cronin effect and the dynamics
of the Color Glass Condensate. 

Particularly, the dilepton transverse momentum distribution presents the
suppression of the Cronin type peak, as observed in the inclusive
observables, if a non-local Gaussian is used. Such behavior is
observed in the Fig. (\ref{rationonloc}). At the LHC energies, at
forward rapidities, the effect of suppression increases
(Fig. (\ref{rationonloc})). Such large suppression at high energies
gives an indication that dilepton transverse momentum provides a clear
probe of the Color Glass Condensate description of the high energy
hadronic interactions in the forward rapidity region.  Moreover, the
dilepton $p_T$ spectra should be used to investigate the properties of
the Cronin effects and indicates it as an initial state effect.  Our
results confirm the studies of Refs. \cite{Raufpp,Rauf2,BMS}
concerning the saturation effects. In addition to this analysis, in a
recent work \cite{FQZ}, the high $p_T$ and low mass region in
the dilepton production in perturbative QCD with all-order resummation
was pointed as a good probe of the gluon distribution, as was
indicated in the Refs. \cite{DYdip} considering the dipole
approach. Also, the mass distributions of the dileptons investigated
in the CGC should identify effects of saturation at small mass region
\cite{dileptonJamal}. The ensemble of these features shows that
dilepton production is an observable that deserves to be measured,
once it carries information about the high density nuclear system.

\section*{Acknowledgments}

We would like to thank L. McLerran, E. Iancu and R. Venugopalan for
discussions during the IX Hadron Physics held in Brazil, K. Safarik
for the elucidating comments on experimental features of the muon
detector at LHC, M. V. T. Machado for useful comments, and F. Gelis
for the comments about Fourier transform of the CGC fit dipole cross
section, for providing a code to evaluate quickly the Fourier transform
integrals and by a careful reading of this manuscript. This work was
supported by CNPq, Brazil.

\end{document}